\documentclass[aip,jcp,preprint]{revtex4-1}
\usepackage{graphicx}
\usepackage{dcolumn}
\usepackage{bm}
\begin{document}
\title{Generalized Coupling Parameter Expansion: Application to Square well and Lennard-Jones fluids} 
\author{A. Sai Venkata Ramana} %
\affiliation{Theoretical Physics Division, Bhabha Atomic Research Centre, Mumbai, 400 085, India}
\date{\today}
\begin{abstract}
The coupling parameter expansion in thermodynamic perturbation theory of simple fluids is generalized to include the 
derivatives of bridge function with respect to coupling parameter. We applied seventh order version of the theory to 
Square-Well (SW) and Lennard-Jones(LJ)fluids using Sarkisov Bridge function. In both cases, the theory reproduced the 
radial distribution functions obtained from integral equation 
theory (IET) and simulations with good accuracy. Also, the method worked inside the liquid-vapor coexistence region where 
the IETs are known to fail.  In the case of SW fluids, 
the use of Carnahan-Starling expression for Helmholtz free energy density of Hard-Sphere reference system has 
improved the liquid-vapor phase diagram(LVPD) over that obtained from IET with the same bridge function. 
The derivatives of the bridge function are seen to have significant effect on the liquid part of the LVPD.
 For extremely narrow SW fliuds, we found that the third order theory is more accurate than the higher order versions.
 However, considering the convergence of the perturbation series, we concluded that the accuracy of the third order
 version is a spurious result.
We also obtained the surface tension for SW fluids of various ranges. Results of present theory 
and simulations are in good agreement.
 In the case of LJ fluids, the equation of state obtained from the present method matched with that 
obtained from IET with negligible deviation. 
We also obtained LVPD of LJ fluid from virial and energy routes and found that there is slight inconsistency between the two 
routes. The  applications lead to the following conclusions. In cases where reference system properties are known accurately,
 the present method gives results which are very much improved over those obtained from the IET with the same bridge 
function.
In cases where reference system data is not available, the method serves as an alternative way of solving the Ornstein-
Zernike  equation with a given closure relation with the advantage that solution can be obtained throughout the phase 
diagram with a proper choice of the reference system.
\end{abstract}

\maketitle 

\section{Introduction}
Describing accurately the thermodynamics and structure of fluids with short ranged inter-particle potentials has been a long standing problem in liquid state theory. Various methods
have been developed to attack the problem. They can be divided into two categories. One category is based on integral equation theory(IET)
and the other is based on thermodynamic perturbation theory(TPT) \cite{hansen}. Methods based on IET take into account correlations between particles and aim at accurate 
description of structural properties of liquids i.e., the radial distribution function (RDF), direct correlation function (DCF) etc.
The IETs involve solving the Ornztein-Zernike equation(OZE) coupled with a closure relation.
 Thermodynamic properties like internal energy and pressure are obtained as integrals of RDF. 
 But inside the two phase region, most of the closures do not have a solution\cite{corn}. Apart from this, the IET method suffers from
 lack of thermodynamic consistency between the energy, virial and compressibility routes.
Various closure relations were proposed to minimize the inconsistency\cite{bomont}. Also closures with adjustable parameters have been proposed to enforce 
thermodynamic consistency by adjusting them. The requirement of consistency makes the computations complicated.
Despite the improvements, accurate description of thermodynamic and structural  properties of simple liquids with narrow potentials is still lacking. 

In earlier formulations of TPT, the structure and thermodynamic properties of reference system are assumed to be known 
and the 
thermodynamic properties of actual system are obtained as a small perturbation over the reference system. The structure of the reference system and the
actual system are assumed to be the same. Within this approach, it is not possible to go beyond second order term in the perturbation series as they require 
higher order correlation functions\cite{barker}. 
 Zhou\cite{zhou1} developed a method in which the Helmholtz free energy of the system is written as a series over coupling parameter $\zeta$.
 This method requires derivatives of the radial distribution function (RDF) at $\zeta = 0$. Zhou calculated the derivatives using finite-difference method and could
 obtain terms in the perturbation series up to sixth order\cite{zhou2}. However, as one goes to higher and higher orders, errors due to
numerical differentiation creep in and the data sets of derivatives of RDF require additional smoothing to remove fluctuations. So
 it is difficult to get derivatives of RDF to arbitrary order. On the other hand, for very short-ranged inter-particle potentials, the
 convergence of the free energy series becomes very slow and requires higher order terms to be evaluated. 

Recently, we addressed the problem in a more fundamental way combining the ideas of IET and TPT. 
We assumed that the RDF and the DCF of a
system can be written as Taylor series in coupling parameter $\zeta$ at $\zeta=0$. We showed that the terms in Taylor series of both RDF and DCF (i.e., their derivatives) 
can be obtained by solving a simple set of equations obtained by differentiating the OZE and the generalized closure
$w.r.t.$ $\zeta$. The formula for free energy given by Zhou can be derived within our formalism apart from the RDF and DCF of the actual 
system. Within this approach we obtained DCF, RDF and LVPDs for Square Well (SW) fluids of various ranges up to seventh order.
By comparing the third, fifth and seventh order versions of our theory, we concluded that the perturbation series for Helmholtz free energy
has practically converged by seventh order. However, the series for RDF and DCF didn't converge for low densities and low temperatures for narrow square well fluids by seventh order.
Even though the Helmholtz free energy series has converged, we found that there was significant deviation of the obtained LVPD using seventh order TPT from simulation 
results for narrow SW fluids. The deviation and slow convergence could be because of two reasons: Firstly, we neglected the derivatives of bridge function $w.r.t.$ $\zeta$ in our 
calculations and secondly the error in the bridge function we have chosen for calculation. 
 
Also, in our previous work, we were able to obtain the RDF and DCF inside the
spinodal and close to the critical region without any problems of convergence of the iteration scheme even
 though our numerical scheme is very simple.
 This raises questions about the existence of the solution to OZE inside the spinodal region.
According to Sarkisov\cite{sarkisov} , solutions to OZE exist in the spinodal region also. However, the IET methods available are not
 convergent in that region. 
  The RDF and DCF in the two-phase region
 find applications in the density functional theory of fluids \cite{evans}. One example is the calculation of surface tension. This calculation requires the DCF
 in the two phase region. As the IETs doesn't have solution in some part of the LVPD, earlier calculations of surface tension were done using an interpolation formula
given by Ebner $et. al.$\cite{ebner} to obtain $c(r)$ inside the two-phase region. 

In present paper, we address these issues using a generalized version of our method with application to SW fluids.
 We use the bridge function proposed by 
Sarkisov\cite{sarkisov} with slight modification for SW fluids\cite{mendoub} and include its derivatives $w.r.t.$ $\zeta$ in the calculation. 
We obtain RDFs, isotherms and LVPDs for SW fluids of various ranges and compare with those obtained from IET and simulations.
We also calculate the surface tension for SW fluids of ranges $1.25$, $1.375$ and $1.75$ using the expression obtained from the square gradient functional for Helmholtz free energy and compare with simulation results.
 As an application of our theory to non-hard sphere reference systems, we apply it to Lennard-Jones (LJ) fluid. 
 We use the Sarkisov bridge function for both reference system and perturbation part.
The RDFs, isotherms and LVPD for LJ fluid are obtained and compared with those obtained from IET and simulations wherever 
available. 
The paper is organized as follows.  In Section II we discuss the method briefly. In Section III we apply the theory to SW fluids to LJ fluids and the results are analyzed. The paper is concluded in Section IV.

\section{ Theory}
Basic theory is the same as explained in Reference[12] except for a generalization to include derivatives of bridge function in the calculation.
Consider a fictitious system at temperature $T$ and density $\rho$ with interaction potential $u(\zeta,r)$ given by
\begin{equation}
u(\zeta, r) = u_{ref}( r) + \zeta u_{pert}(r) \label{u-zeta}
\end{equation}
$\zeta$ is a coupling parameter. When $\zeta=0$, the fictitious system becomes reference system with interaction potential $u_{ref}(r)$. A non-zero $\zeta$ will add a perturbation to $u_{ref}(r)$ as shown in the above equation. $\zeta=1$ corresponds to
the potential of the system under consideration. 
 We postulated that the RDF and DCF of the system with potential $u(\zeta, r) $ can be written as a McLaurin series in $\zeta$, that is,
\begin{equation}
c(\zeta,r) = c(0,r) + \zeta\left(\frac{\partial c}{\partial \zeta}\right)_{\zeta = 0} + \frac{\zeta^2}{2!}\left(\frac{\partial^2c}{\partial \zeta^2}\right)_{\zeta = 0} + .........  \label{c-series}
\end{equation}
\begin{equation}
g(\zeta,r) = g(0,r) + \zeta\left(\frac{\partial g}{\partial \zeta}\right)_{\zeta = 0} + \frac{\zeta^2}{2!}\left(\frac{\partial^2g}{\partial \zeta^2}\right)_{\zeta = 0} + .........  \label{g-series}
\end{equation}
\noindent
 Hereafter we shall denote $n^{th}$ partial derivative of any function $X(\zeta, r)$ $w.r.t\; \zeta$ as $X^{(n)}(\zeta,r)$.  
\noindent
The radial distribution function can be written as 
\begin{eqnarray}
g(\zeta,r) &=& \exp(\phi(\zeta,r) ) \label{g-def-1}  \nonumber \\  
\phi(\zeta,r) &=& -\beta \left( u_{ref}(r) + \zeta u_{pert}(r)\right) + y(\zeta,r) + B(\zeta,r)  \label{gg}  
\end{eqnarray}
where $y(\zeta, r)$ is the indirect correlation function defined as $h(\zeta, r) - c(\zeta, r)$ and $h(\zeta, r) = g(\zeta, r) -1$ is the total 
correlation function of the fictitious fluid. $-\phi(\zeta,r)/\beta$ is called the potential of mean force. The expression for $\phi(\zeta,r)$ in 
Eq.(\ref{gg}) is exact\cite{hansen}.
$B(\zeta, r)$ called the bridge function \cite{bomont}. Since $g(\zeta, r)$, and hence $h(\zeta, r)$, as well as $c(\zeta, r)$ are expanded in a series in $\zeta$, the correlation function $y(\zeta, r)$ 
is also a series in $\zeta$. The $n^{th}$ order coefficient in its series is given by  $y^{(n)}(\zeta, r) = h^{(n)}(\zeta, r) - c^{(n)}(\zeta, r)$. Generally, $B(\zeta,r )$ is chosen as a function of $y(\zeta,r)$.
Several approximations to $B(\zeta, r)$ in terms of $y(\zeta, r)$ are available \cite{bomont} in literature.


The general expression for $n^{th}$ order derivative of $g(\zeta,r)$ from Eq.(\ref{gg}) is
\begin{equation}
g^{(n)}(\zeta,r)= \sum_{m=0}^{(n-1)} [C_m^{(n-1)}]\; \phi^{(n-m)}(\zeta,r)\; g^{(m)}(\zeta,r), \; n \geq \;1 \label{g-def-n}  
\end{equation}
\noindent
where $[C_m^{(n-1)}]$ is the binomial coefficient. The derivatives $\phi^{(n)}(\zeta,r)$ are given by 
\begin{equation}
\phi^{(n)}(\zeta,r) =  -\beta  u_{pert}(r)\; \delta_{n,1} + y^{(n)}(\zeta,r)+ B^{(n)}(\zeta,r),  \;  n \geq 1 \label{f-der-n}  
\end{equation}
\noindent
where $\delta_{n,1}$ is the Kronecker delta. The derivatives  $g^{(n)}(\zeta,r)$  can be computed 
 using Eq.(\ref{g-def-n}) in a recursive manner, using $\phi^{(n)}(\zeta,r)$ either from initial guess or previous iteration.

\noindent
Since
\begin{equation}
c(\zeta,r) = g(\zeta,r) - y(\zeta,r) -1,  \label{c-closure-1}
\end{equation}
\noindent
 its $n^{th}$ order derivative is
\begin{equation}
c^{(n)}(\zeta,r)= g^{(n)}(\zeta,r) - y^{(n)}(\zeta,r),\; n \geq 1.    \label{c-closure-n}
\end{equation}

To get another set of relations between $c^{(n)}(\zeta,r)$ and $y^{(n)}(\zeta,r)$, we consider the Ornstein Zernike Equation (OZE) in Fourier space:
\begin{equation}
h(\zeta,k) = \frac{c(\zeta,k)}{1 - \rho c{(\zeta,k)}}  \label{oze} 
\end{equation}
\noindent
where $h(\zeta,k)$ and $c(\zeta,k)$ are the Fourier transforms of $h(\zeta,r)$ and $c(\zeta,r)$, respectively. 

Also, structure factor $s(\zeta,k)$ is defined as
\begin{equation}
s(\zeta,k) = \frac{1}{1-\rho c(\zeta,k) }  \label{sfac}
\end{equation}
Therefore, using above equation in OZE i.e., Eq.(\ref{oze}) we get
\begin{equation}
s(\zeta,k) = 1 + \rho h(\zeta,k) \label{sh}
\end{equation} 
Now, differentiating Eq.(\ref{sfac}) $n$-times, we get
\begin{equation}
s^{(n)}(\zeta,k) = [ s^{(0)}(\zeta,k) ]\; \rho\; \sum_{m=0}^{(n-1)} [C_m^n]\; c^{(n-m)}(\zeta,k)\; s^{(m)}(\zeta,k), \; n \geq 1  \label{sfac-n-1} 
\end{equation}
Using Eq.(\ref{sh}) the derivatives of $ y(\zeta,k)=h(\zeta,k) - c(\zeta,k)$ are expressed as
\begin{equation}
y^{(n)}(\zeta,k) =  \rho^{-1}\; s^{(n)}(\zeta,k) - c^{(n)}(\zeta,k), \; n \geq 1  \label{yk-n} 
\end{equation}
Thus in order to obtain up to $N^{th}$ order term in Taylor series expansion of both $c(1,r)$ and $g(1,r)$, $N$ coupled linear 
equations in real-space and $N$ coupled linear equations in Fourier space have to be solved simultaneously with $\zeta=0$.

For example, to obtain Taylor series expansion up to second term, the set of four equations given by
\begin{eqnarray}
c^{(1)}(0,r) &=& ( -\beta u_{pert}(r) +y^{(1)} (0,r)+B^{(1)} (0,r) ) g(0,r) - y^{(1)} (0,r)  \label{c10} \\
c^{(2)}(0,r) &=& ( -\beta u_{pert}(r) +y^{(1)}(0,r)+B^{(1)} (0,r) )^2 g(0,r)  \\ \nonumber
                  &  & + B^{(2)}(0,r) g(0,r)+ y^{(2)}(0,r) (g(0,r)-1)  \label{c20} \\
y^{(1)}(0,k) &=& c^{(1)}(0,k) (s^2 (0,k) -1)  \label{yk10} \\
y^{(2)}(0,k) &=& c^{(2)}(0,k) (s^2(0,k) - 1) + 2 \rho (c^{(1)}(0,k))^2 s^3(0,k)  \label{yk20}
\end{eqnarray}
have to be solved (where $g(0,r)$ is $y(0,r) + c(0,r) + 1$ ).

The CPE for Helmholtz free energy density 
$f(T,\rho)$ of a homogeneous fluid at a given temperature $T$ is given by \cite{hansen}
\begin{equation}
f(T,\rho) = f_{ref}(\rho) +  \frac{\rho^2}{2} \int_0^1 d\zeta \int d\vec r\; u_{pert}(r)g(\zeta,r) \label{f}
\end{equation}
\noindent
where $f_{ref}(\rho)$ is the free energy density of the reference system.
For notational simplicity we do not show temperature dependence of $f(T,\rho)$ explicitly hereafter.
 Substituting Eq.(\ref{g-series}) in Eq.(\ref{f}) and integrating over $\zeta$, we get
\begin{equation}
f(\rho) = f_{ref}(\rho) +  \frac{\rho^2}{2} \int d\vec r\; u_{pert}(r)\left(g(0,r) + \frac{1}{2!}\; g^{(1)}(0,r) + \frac{1}{3!}\; g^{(2)}(0,r) + ........ \right) \label{f1}
\end{equation}
\noindent
Here we have used the shortened notation for the derivatives $\left( \partial ^n g(r)/\partial \zeta^n \right)_{\zeta =0}$ which is readily obtained as $y^{(n)}(0,r) + c^{(n)}(0,r)$.
  Thus the method provides the 
DCF, RDF as well as the free energy density. The Taylor series of RDF truncated up to second order  gives third order
 CPE for $f(\rho)$. The numerical procedure we used is as explained in our previous paper\cite{sai} and is not repeated here. Eq.(\ref{f1}) has been first derived by Zhou\cite{zhou1} in a different way.

\section{Applications}
\subsection{Square Well Fluids}
Seventh order version of above mentioned theory is applied to SW fluids. 
We used the closure proposed by Sarkisov with slight modification by Mendoub\cite{mendoub}.
\begin {equation}
B(\zeta,r) = (1 + 2y^*(\zeta,r))^{1/2} - 1 - y^*(\zeta,r) \label{b-m}
\end{equation}
where 
\begin{eqnarray}
y^*(\zeta,r) &=& y(\zeta,r) + \rho f_M(\sigma^+)/2,\mbox{    $ r < \sigma$}  \\ \nonumber
		     &=& y(\zeta,r) + \rho f_M(r)/2,\mbox{                $ r \ge \sigma$}
\end{eqnarray}
and $\sigma$ is the Hard sphere diameter. We obtained the RDF and DCF of the reference system using the same bridge function by solving the OZE using a similar procedure as explained above to maintain consistency.

\begin{figure}
\includegraphics[scale=0.5,angle=-90]{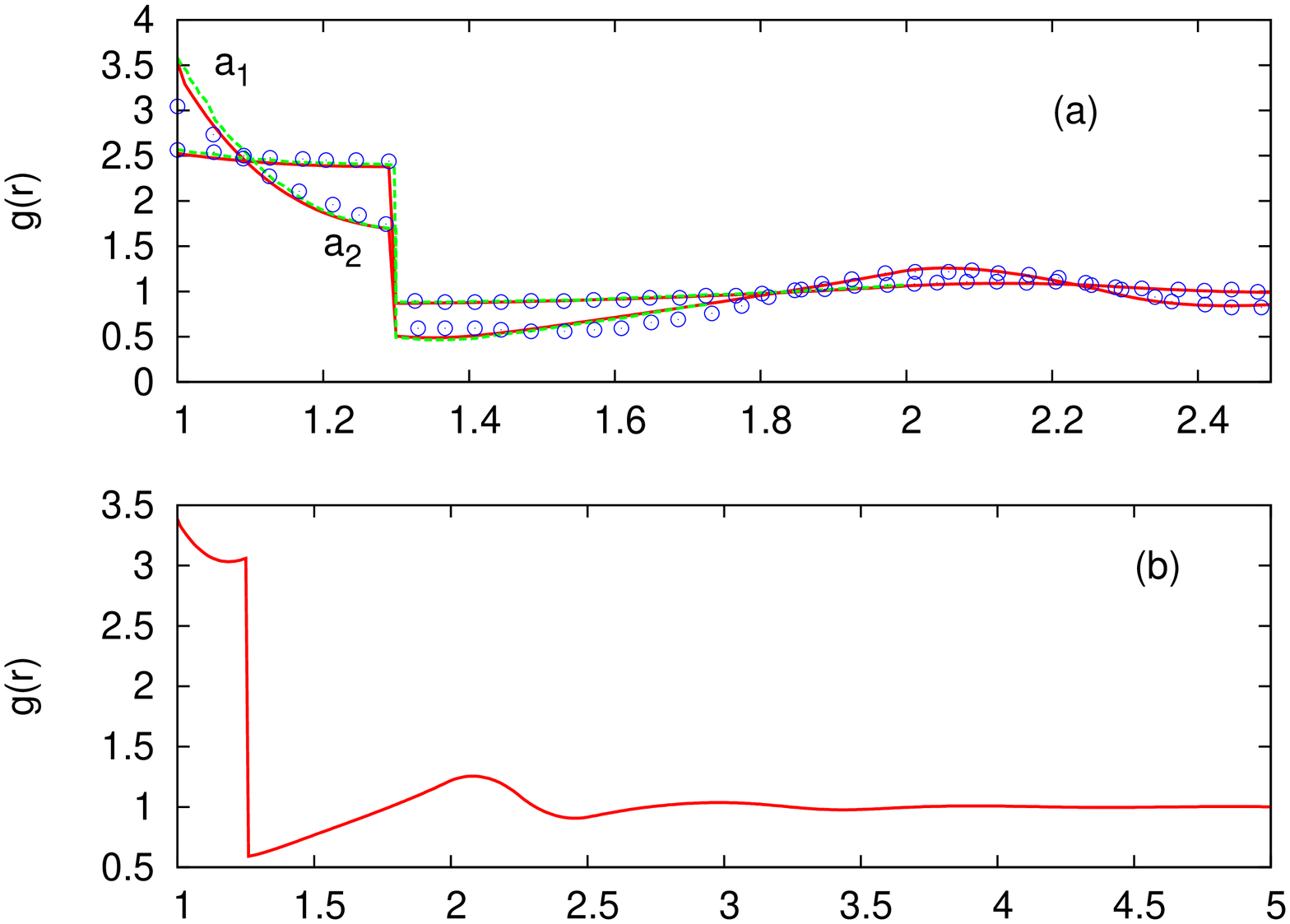}%
 \caption{\label{1} (Color Online) (a) $g(r)$ for SW fluid of range 1.3 for densities 0.2(curve $a_2$),0.8(curve $a_1$) and $T=1.0$. (Circles: simulation results\cite{largo}); (Dashes: IET results \cite{mendoub}); (Solid lines: results from present calculations). (b) $g(r)$ for  SW fluid of width 1.25, temperature $T =0.65$ and density $0.4$.} 
 \end{figure}

 Reduced units ($\epsilon/k_B = \sigma =1$, where $\epsilon$ is the well depth and $\sigma$ is hard 
sphere diameter) are used throughout the paper.
For convenience, we denote $c(0,r)$, $g(0,r)$ by $c_0(r)$, $g_0(r)$ respectively and 
$c(1,r)$, $g(1,r)$ by $c(r)$, $g(r)$ respectively.
  In Fig.(\ref{1}) we compare $g(r)$ obtained using $7^{th}$ order version of our TPT and that obtained 
 through IET\cite{mendoub} for SW fluid of range $1.3$ for densities $0.2$ and $0.8$ and temperature $T=1.0$.
Clearly, there is negligible difference between results obtained using present method and those obtained from IET.
Also, the agreement with simulation results is good. We observed that except for very low temperatures and low densities,
   results of fifth order and seventh order TPTs have negligible deviation showing that
 the Taylor series has converged. Convergence of the iteration scheme has been good in the whole 
phase diagram. $g(r)$ obtained by our method for SW fluid of range $1.25$ in the spinodal region 
at $\rho = 0.4$ and $T=0.65$ is shown in Fig.(\ref{1}).
Above observations show that the present method may be viewed as an alternative way of solving the 
OZE. 
 
\begin{figure}
\includegraphics [scale=0.5,angle=-90]{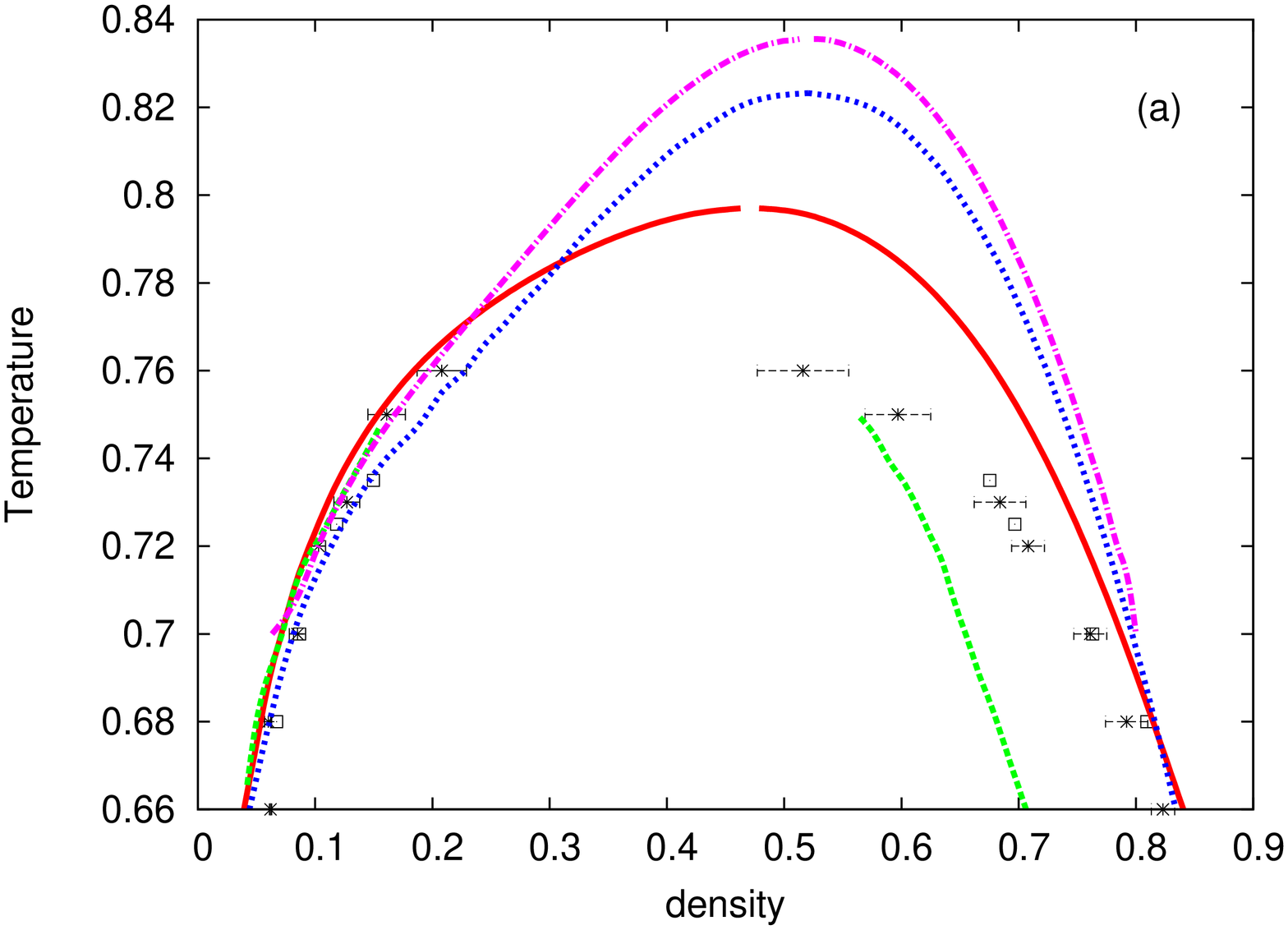}%

 \includegraphics [scale=0.5,angle=-90]{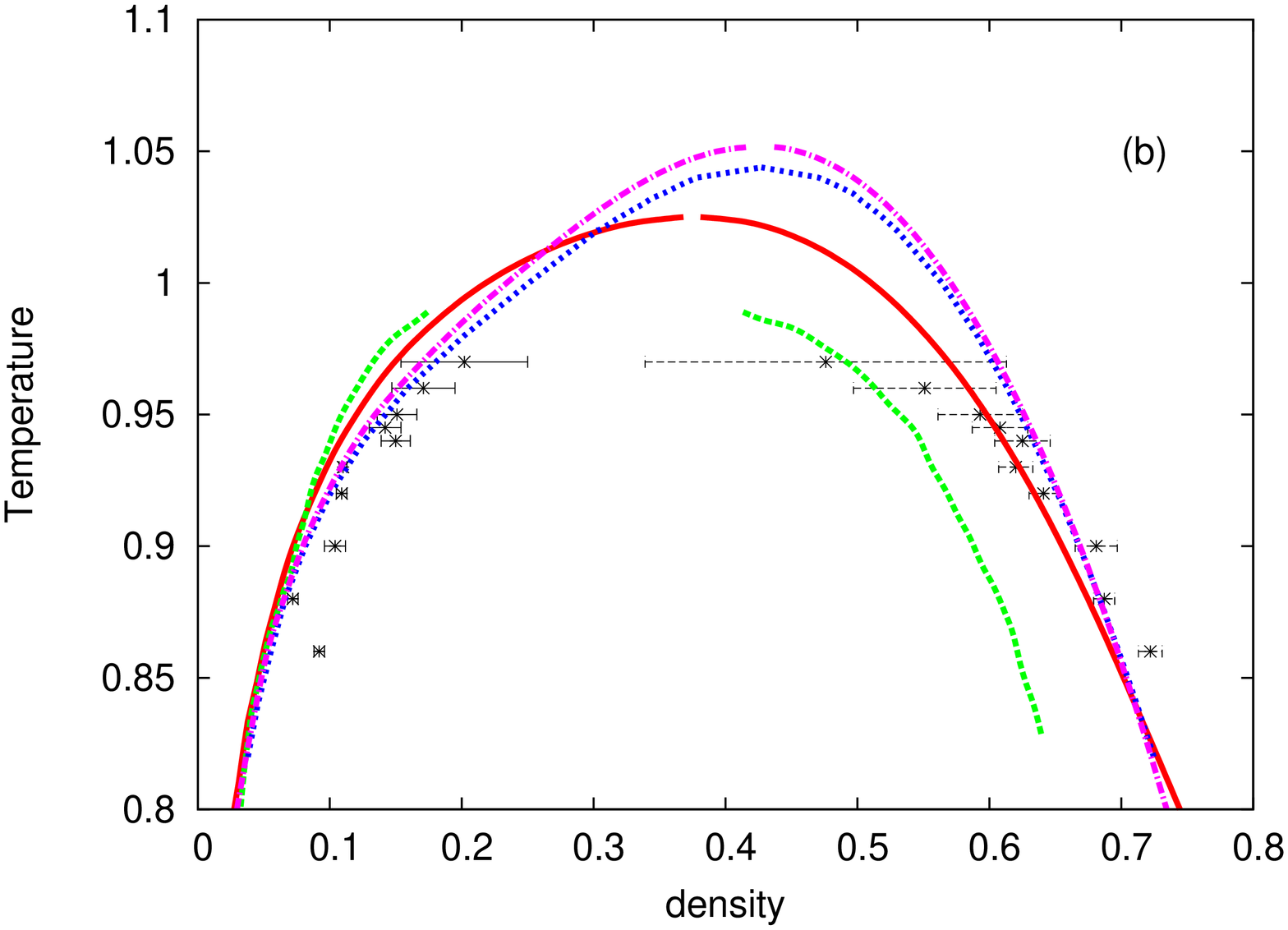}
 \caption{\label{2} (Color Online)  LVPD of SW fluid of range 1.25(panel (a)) and LVPD of SW fluid of range 1.375(panel (b)). ( dotted line: 7th order TPT with Malijevsky Labik bridge function\cite{sai}); (solid line: 7th order TPT with Sarkisov B(r) including derivatives of B(r)); (dash-dot: 7th order TPT with Sarkisov B(r) without including derivatives of B(r)); (Short dashes: IET results with Sarkisov B(r) obtained from Mendoub et. al.\cite{mendoub});( Squares and stars are simulation results\cite{vega,rio}); } 
 \end{figure}

In Fig.(\ref{2}) we give plots of LVPD for SW fluids of ranges $1.25$ and $1.375$
respectively obtained using seventh order version of TPT with and without including the derivatives of bridge function 
in the calculations. We used the Carnahan-Starling(CS) expression for Helmholtz free energy density of the reference system. Our results are compared with simulations, those obtained from IET\cite{mendoub} 
and our previous results using Malijevsky-Labik brdige function neglecting the derivatives of bridge function.
Results of the present calculations are very much improved in the liquid part of the LVPD over those of Mendoub
\cite{mendoub}. The main reason for this improvement is the use of the CS expression for free energy density.
A comparison of the LVPDs obtained by including derivatives of the bridge function in the calculation and those 
obtained by neglecting them shows that the correction due to the derivatives becomes important close to the critical region and along the liquid side of the phase diagram. Neglecting them leads to shifting of the critical point to high density region.

\begin{figure}
\includegraphics[scale=0.5,angle=-90]{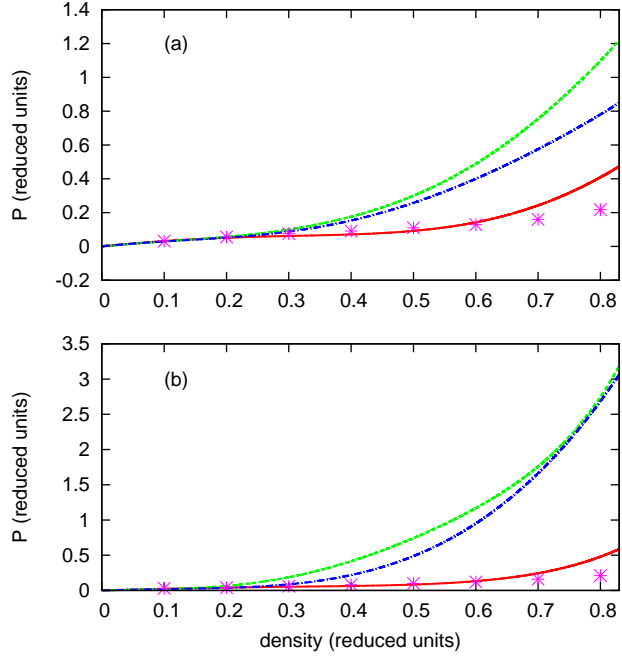}
 \caption{\label{3} (Color Online)  :(a) Pressure(P) of SW fluid of range 1.04 at $T = 0.37$.(b) P of SW fluid of range 1.01 at $T= 0.255$. P and T are in reduced units. Stars: Simualtion results \cite{zhou3}. (Solid Line: $3^{rd}$ order TPT), (Dash-dot: $5^{th}$ order TPT), (Dash: $7^{th}$ order TPT)}
 \end{figure}

In Fig.(\ref{3}), pressure isotherms obtained using $3^{rd}$, $5^{th}$ and $7^{th}$ order TPT for SW fluids of 
extremely short ranges $1.01$ and $1.04$ are plotted for temperatures $0.255$ and $0.37$ respectively. The figures
depict that the $3^{rd}$ order results are much closer to the simulation results. However, a comparison of the 
 three versions shows that the series is indeed convergent at each density point. But it is converging to a point
 away from the simulation results. This is because of the inaccuracy of the bridge function for very short ranged 
 fluids. Thus the apparent closeness of the third order TPT results to the simulation results is spurious.

Possibility of obtaining $c(r)$ in two phase region opens up applications in density functional theory(DFT) of liquids.
One simple application is calculation of surface tension of liquids. We use the $c(r)$ obtained from
seventh order version of TPT to obtain surface tension for SW fluids. Formula for surface tension 
is obtained by Yang et. al. \cite{yang}  from 
square-gradient functional for Helmholtz free energy of inhomogeneous fluids. A brief derivation is as follows:
 The square-gradient approximation for Helmholtz free energy functional of an inhomogeneous liquid occupying volume $\Omega$ at temperature $T$ is given by

\begin{equation}
F[\rho(\vec r)] = \int d\vec r\ \{ f(\rho(\vec r)) + f_g(\rho(\vec r))|\nabla \rho(\vec r)|^2 \} \label{sg-f}
\end{equation}

\noindent
where $f(\rho)$ is Helmholtz free energy density of homogeneous liquid. 
 The second term is the effect of inhomogeneity.
$\rho(\vec r)$ is the number density in an infinitesimal volume around $\vec r$.$F[\rho(\vec r)]$, $f(\rho(\vec r))$ and 
$f_g(\rho(\vec r))$ are all functions of $T$ even though the dependence is not shown explicitly.

The coefficient $f_g$ of gradient term also referred as influence parameter is
\begin{equation}
f_g [\rho(\vec r)] = \frac{1}{12\beta}\int d\vec r' r'^2 c[ \rho(\vec r),  r' ]   \label{f-G-rho-vec}
\end{equation}
We assume that liquid-vapor interface is flat and that $z$-axis is normal to the interface pointing out into the vapor from the
liquid. In such a case, Eq.(\ref{sg-f}) becomes

\begin{equation}
F[\rho(z)] = A\int dz \ \{ f(\rho(z)) + f_g(\rho(z))\left |\frac {d\rho(\vec r)}{dz}\right |^2 \} \label{sg-f2}
\end{equation}
where $A$ is the surface area. Grand free energy of the system is given as
\begin{equation}
\Gamma[\rho(z)] = F[\rho(z)] - \mu N \label{gp}
\end{equation}
where $\mu$ is the chemical potential of the system and $N$ is the total number of particles

Minimizing $\Gamma$ $w.r.t.$ $\rho$ we get
\begin{equation}
\frac{d}{dz}\left[f_g(\rho) \left( \frac{d\rho}{dz} \right)^2 \right] = \frac{d\gamma(\rho)}{dz} \label{v}
\end{equation}
where $\gamma(\rho)$ is the grand free energy density. 

Integrating above Eq.(\ref{v}) with boundary conditions 
\begin{equation}
\rho(z\to \infty) = \rho_g;\rho(z\to -\infty) = \rho_l \mbox{    }and \mbox{    } \frac{d\rho(z\to \pm \infty)}{dz} \label{bc}
\end{equation}
gives
\begin{equation}
\frac{d\rho}{dz} = \left[\frac{\gamma(\rho) - \gamma(\rho_l)}{f_g(\rho)} \right]^{1/2} \label{drdz}
\end{equation}
where $\rho_l$ and $\rho_g$ are coexisting liquid and vapor densities for temperature $T$ under consideration.
Surface tension $S$ can be calculated using the formula 
\begin{equation}
S = 2 \int_{-\infty} ^{\infty} f_g(\rho) \left( \frac{d\rho}{dz} \right)^2 dz \label{s}
\end{equation}
Using Eq.(\ref{drdz}) in above equation gives
\begin{equation}
S = 2\int_{\rho_g}^{\rho_l} [f_g(\rho)(\gamma(\rho) - \gamma(\rho_l))]^2 \label{s2}
\end{equation}
In the above equations $\gamma(\rho_l)$ may be replaced by $\gamma(\rho_g)$ also as both have same value.

\begin{figure}
\includegraphics [scale=0.5,angle=-90]{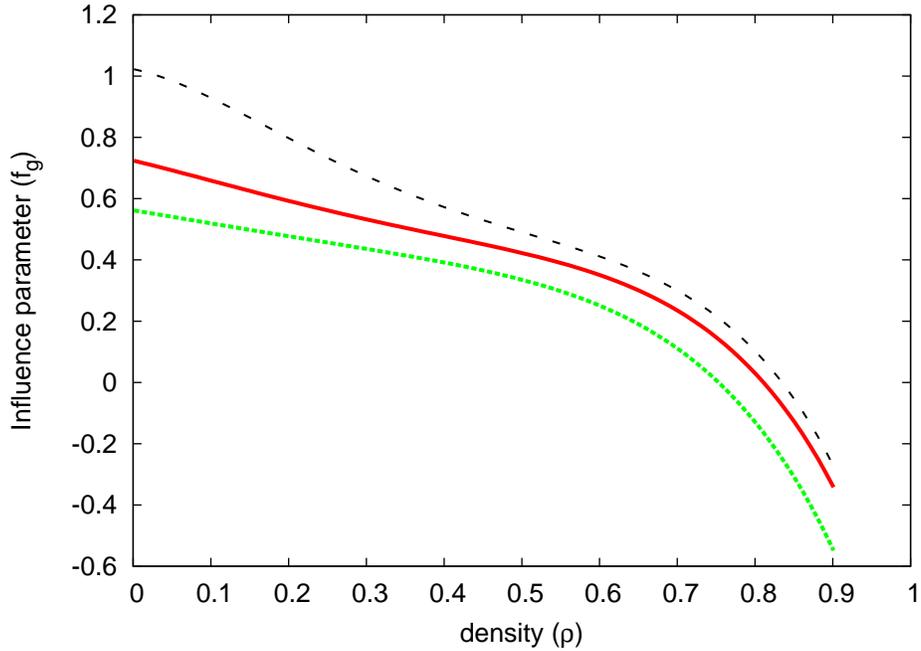}  
 \caption{\label{4}(Color Online) Coefficient of square gradient functional for SW fluid of range 1.25. Curves from top to bottom are for temperatures T = 0.6, 0.8 and 1.0.}
 \end{figure}

\begin{figure}
\includegraphics [scale=0.5,angle=-90]{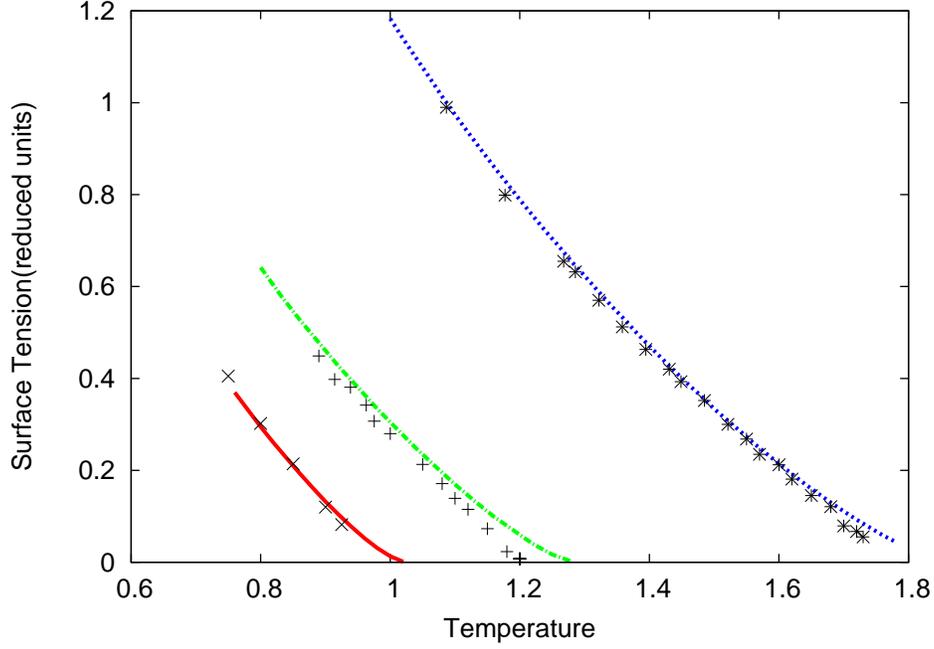}  
 \caption{\label{5}(Color Online) Surface tension for SW fluids in reduced units. From left to right are for ranges 1.375, 1.5 and 1.75. stars, pluses, crosses are simulation results\cite{singh,lopez}. Lines: $7^{th}$ order TPT.}
 \end{figure}

Above explained formalism is applied to SW fluids of ranges $1.375$, $1.5$ and $1.75$ with $c(r)$ obtained from seventh order
 TPT as explained above. Fig.(\ref{4}) shows $f_g(\rho)$ as a function of $\rho$ for different temperatures for SW fluid of 
range $1.25$. From Fig.(\ref{4}) it can be seen that $f_g(\rho)$ becomes negative at high densities which is unphysical. This 
might be an artifact of the approximate bridge function used. However, within the binodal, where $f_g$ is required for the 
calculation of surface tension(S), it is positive. Surface tensions obtained from our calculation are plotted as a function of $T$
 in Fig.(\ref{5}). Our results compare well with simulations except close to the critical temperatures. 
\subsection{Lennard Jones Fluid}
\begin{figure}
\includegraphics[scale=0.5,angle=-90]{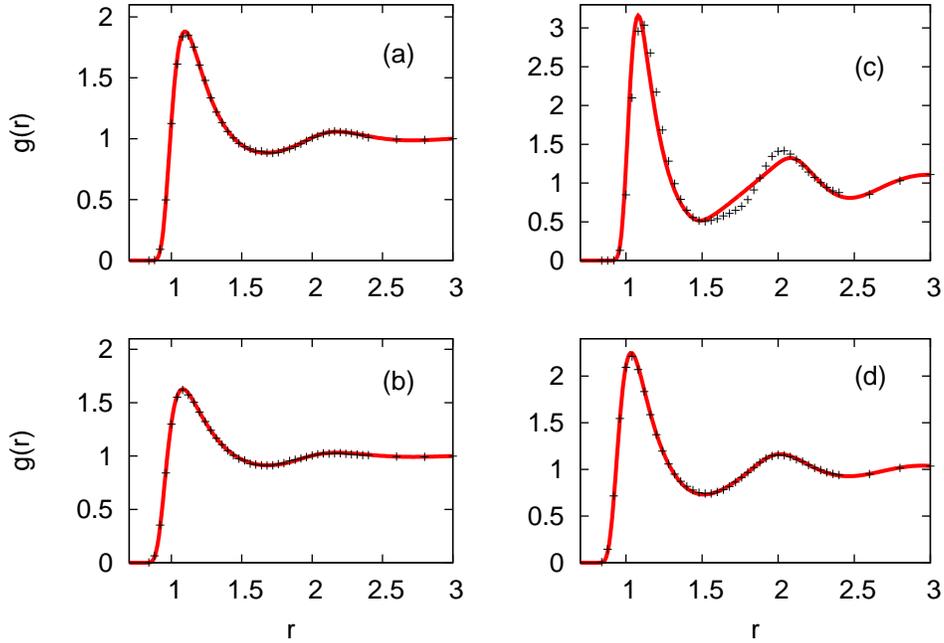}
 \caption{\label{6} (Color Online) LJ rdfs for different temperatures and densities obtained using $7^{th}$ order TPT. (pluses: Simulation results\cite{verlet}),(Top left: $T=1.552, \rho = 0.45$),(Bottom left: $T=2.934, \rho = 0.45$),(Top right: $T=0.658,\rho=0.85$),(Bottom right: $T = 2.888,\rho = 0.85$). $T$ and $\rho$ are in reduced units}
 \end{figure}

\begin{figure}
\includegraphics[scale=0.5,angle=-90]{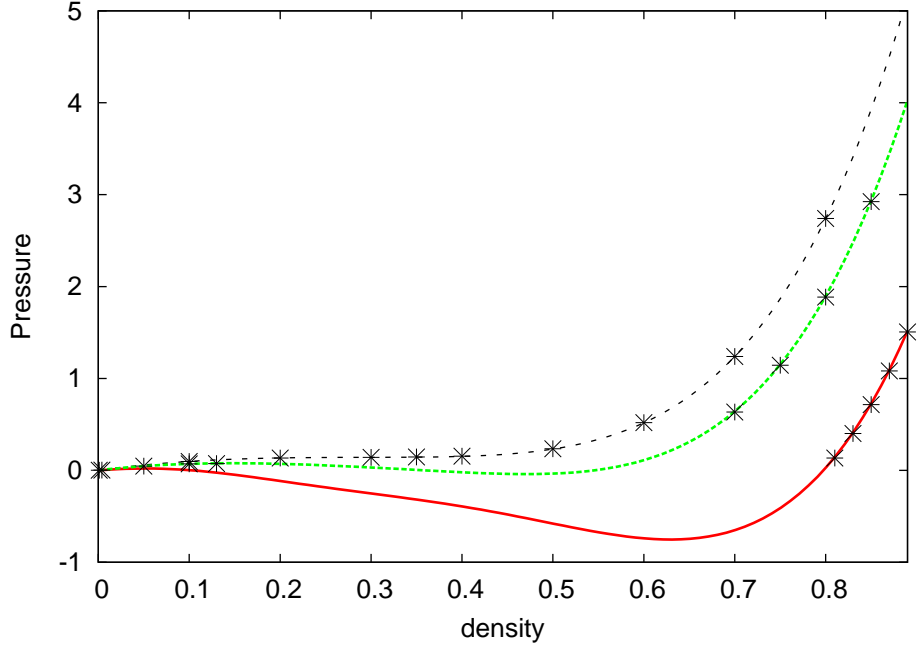}
 \caption{\label{7} (Color Online)  : Pressure (reduced units) of LJ fluid for different temperatures. Stars: Values obtained using IET by Sarkisov\cite{sarkisov}. Lines: 7th order TPT results with Sarkisov bridge function. (Solid line: T = 0.75),(Dotted line: T = 1.15),(Dashes: T = 1.35)} 
 \end{figure}

\begin{figure}
\includegraphics[scale=0.5,angle=-90]{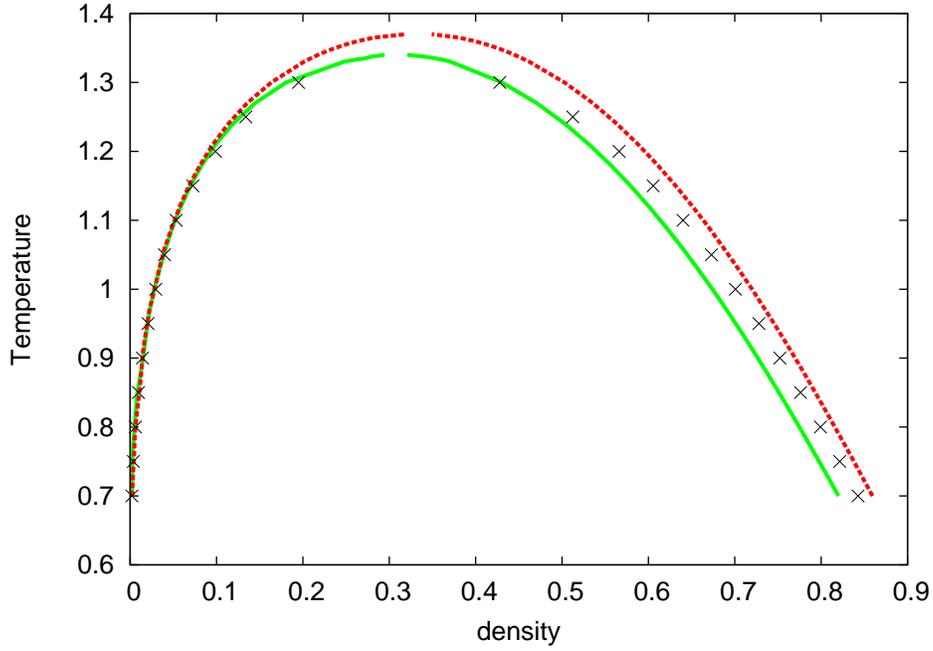}
 \caption{\label{8} (Color Online) LVPD of LJ fluid. (Dashed line: EOS calculated using energy route). (Solid line:EOS obtained using virial route). In both cases Sarkisov bridge function and $7^{th}$ order TPT are used. Crosses are simulation results\cite{lofti}.}
 \end{figure}

The theoretical formalism explained in section II is general and can be applied to non-hard sphere reference systems also.
As an example, we applied the theory to Lennard Jones(LJ) fluid. The Lennard Jones potential was
split into reference and perturbation parts according to the Weeks Chandler and Anderson (WCA)\cite{WCA} method. 
 In this case, we used for $B(\zeta,r)$, the expression proposed by Sarkisov \cite{sarkisov} which is same as Eq.(\ref{b-m}) with
\begin{eqnarray}
y^*(\zeta,r) &=& y(\zeta,r) - \rho \beta u(r_m),\mbox{                     $ r < r_m$}  \\ \nonumber
		     &=& y(\zeta,r) - \rho \beta u(r),\mbox{                                 $ r \ge r_m$}
\end{eqnarray}
where $r_m$ is the minimum energy point of the Lennard-Jones potential. The $c_0(r)$ and $g_0(r)$ of the reference system are obtained
by solving the OZE with the same bridge function. Thus, in effect, we solved the OZE for LJ fluid with the above bridge function 
using the perturbation method. In Fig.(\ref{6}),  we compare the $g(r)$ of LJ fluid obtained
 from seventh order TPT with simulation results.
 There is excellent agreement between the theory and simulation results for the cases shown except 
for a slight deviation in the case of lowest temperature. In Fig.(\ref{7}), Equation of State(EOS) of LJ fluid for various 
temperatures obtained using seventh order TPT is compared with those obtained by solving OZE by Sarkisov\cite{sarkisov}.
Pressure(P) is obtained using the virial formula given by
\begin{equation}
P = \rho k_B T - \frac{1}{6}\rho^2\int_0^{\infty}\frac{du(r)}{dr}g(r) 4\pi r^2 dr \label{Pvir}
\end{equation}
Values obtained by our method matched with negligible deviation from those given by Sarkisov. In this case also, using our method we 
could obtain the pressure for all density points in the phase diagram, as $g(r)$ at any point in the phase diagram could be 
calculated. 
 Again, even close to the critical region, the same 
numerical procedure was sufficient whereas IETs face numerical convergence problems in this region. In Fig.(\ref{8}), LVPD
 of LJ fluid is shown and compared with simulation results. We plotted the LVPD obtained in two ways.  One is obtained 
 by Maxwell construction of pressure isotherm obtained using Eq.(\ref{Pvir}) which is the so called virial route.
Alternatively, LVPD is also obtained using the energy route. It is obtained as 
follows: Pressure isotherm of the reference fluid is obtained using Eq.(\ref{Pvir}) with $u_{ref}(r)$. From this, Helmholtz free 
energy density of the reference fluid is obtained from the formula below
\begin{equation}
f_{ref}(\rho) = \rho k_B T\int_0^{\rho}(\frac{P_{ref}(\rho')}{ \rho'^2} - 1) + \rho k_B T ln(\rho)  \label{fref}
\end{equation}
Once the reference free energy is obtained, method described in section II can be applied to get the free energy density of
the required system. The pressure isotherm is obtained by differentiating the Helmholtz free energy $w.r.t.$ volume of the system.
Maxwell construction is done to get the coexistence points. From the figure, it can be seen that the LVPD obtained from
 both the routes differ slightly along the liquid part of coexistence curve. Also, there is some deviation of both LVPDs from
the simulation results. This is due to the bridge function used. Even though Sarkisov bridge function is supposed to be
quiet accurate, there is some inconsistency between various thermodynamic routes. Imposition of the thermodynamic consistency between 
various routes as a constraint would solve the problem and may improve the accuracy of the results. This can be done in a straight-forward way within the theory presented in the paper.
\section{Conclusion}
We have generalized the TPT of fluids based on coupling parameter expansion to include the derivatives of the bridge function.
The applications presented in the paper let us conclude the following points.(i) Inclusion of derivatives of the bridge function
in the TPT has a significant effect on structural and thermodynamic properties of liquids.
(ii) If accurate reference system data is available, it can be used in the TPT to improve the accuracy of the results
 over those of IET. (iii) The present method may be viewed as an alternative way of solving the OZE with the advantage that solution 
 can be obtained in the whole phase diagram.

\section{Acknowledgements}
I am indebted to Dr. S.V. G. Menon for introducing me to the subject when he was in Theoretical Physics Division, B.A.R.C. I thank Dr. D. M. Gaitonde for useful comments and discussions.

\newpage

\end{document}